\documentclass[aps,pra,groupaddress,twocolumn,floatfix]{revtex4-1}
\usepackage{units}
\usepackage{amsmath}
\usepackage{amssymb}
\usepackage{graphicx}
\usepackage{bm}
\usepackage{multirow,color,relsize,ulem,microtype}

\newcommand{\be}{\begin{equation}}
\newcommand{\ee}{\end{equation}}
\newcommand{\e}{\omega}

\begin{document}

\title{Origin of robust exceptional points : a restricted bulk zero mode}

\author{Jose D. H. Rivero}
\affiliation{\textls[-18]{Department of Physics and Astronomy, College of Staten Island, CUNY, Staten Island, NY 10314, USA}}
\affiliation{The Graduate Center, CUNY, New York, NY 10016, USA}

\author{Li Ge}
\email{li.ge@csi.cuny.edu}
\affiliation{\textls[-18]{Department of Physics and Astronomy, College of Staten Island, CUNY, Staten Island, NY 10314, USA}}
\affiliation{The Graduate Center, CUNY, New York, NY 10016, USA}

\date{\today}

\begin{abstract}
Recently a type of robust exceptional points was found that is insensitive to the coupling disorder in the bulk. Here we show that a disparity emerges when the number of coupled cavities in this one-dimensional array changes from even to odd. The robust exceptional point only exists in the former case, whereas the location of the exceptional point in the latter depends inversely on the size of the cavity array and is subjected to coupling disorder in the bulk. We further show that the exceptional points in these two cases are second and third order, respectively. We elucidate the origin of the robust EP as a restricted bulk zero mode, which shares the same robustness against coupling disorder and has a finite amplitude adjacent to the boundary. This finding enables us to identify robust EPs in higher dimensional systems reliably, which can exist in the presence of either sublattice symmetry or the non-Hermitian particle-hole symmetry evolved from it.
\end{abstract}

\maketitle

\section{Introduction}

Sublattice symmetry plays an important role in the study of topological phases of matter \cite{Hasan,Qi,Alicea,Beenakker,Sarma_RMP,Sarma_QI,Alicea_PRX}. As an example of chiral symmetry, it warrants that the energy spectrum of the system is symmetric about a well-defined energy level, such as the Fermi level or the energy of an uncoupled orbital, which is chosen as the zero energy. Recently, the exploration of sublattice symmetry in non-Hermitian systems \cite{NPreview,NPhyreview,RMP,Moiseyev_book} has also attracted fast growing interest, especially with the advent of topological photonics \cite{Lu} and lasers \cite{St-Jean,Bahari,Bandres,Zhao,Pan,Parto,Leykam}.

One interesting finding indicates that sublattice symmetry evolves into a non-Hermitian particle-hole (NHPH) symmetry \cite{Kohmoto} when imaginary on-site potentials are imposed on a tight-binding model with two sublattices and real couplings \cite{zeromodeLaser}. Different from sublattice symmetry, NHPH symmetry warrants a complex spectrum symmetric about the imaginary axis of the complex energy plane, i.e., $\e_\mu = -\e_\nu^*$. This NHPH symmetry was hidden in many previously studied non-Hermitian systems, including parity-time (PT) symmetric systems \cite{Bender1,PTLaserReview}. Notably, sublattice symmetry can also persist in non-Hermitian systems by allowing the couplings to be asymmetric \cite{Malzard,zeromodeLaser}. In a recent work \cite{REP}, this non-Hermitian sublattice symmetry was applied to a one-dimensional (1D) cavity array, and it was found that the exceptional point (EP), a non-Hermitian degeneracy with identical wave functions \cite{EP1,EP2,EP3,EP4,EP5,EP6,EP_CMT,EP_ring,PT_PRX14}, does not depend on the system size or the coupling disorder in the bulk.

In this work, we first point out that this property holds only when there are an even number of cavities in the array. If there are an odd number of cavities instead, the location of the EP depends inversely on the system size and is subjected to the change of couplings in the bulk. We also reveal that these EPs are third order, in contrast to the robust one that is second order. These findings are included in Sec. II.

Furthermore, we report the persistence of a robust EP when sublattice symmetry is broken, as well as the absence of a robust EP when the system does have sublattice symmetry. These findings enable us to pinpoint the origin of a robust EP: it is not due to sublattice itself but rather to the existence of a \textit{restricted} bulk zero mode, which shares the same robustness against coupling disorder \textit{and} has a finite amplitude adjacent to the boundary. Such a zero mode can exist in the presence of either sublattice symmetry or NHPH symmetry, and in the latter case, additional robustness of the EP exists against variations of imaginary on-site potentials. These results are included in Sec. III, where illustrative two-dimensional (2D) systems are exemplified. Finally, some concluding remarks are given in Sec. IV.

\section{Robust EP}
\label{sec:REP}
\begin{figure}[t]
\includegraphics[clip,width=\linewidth]{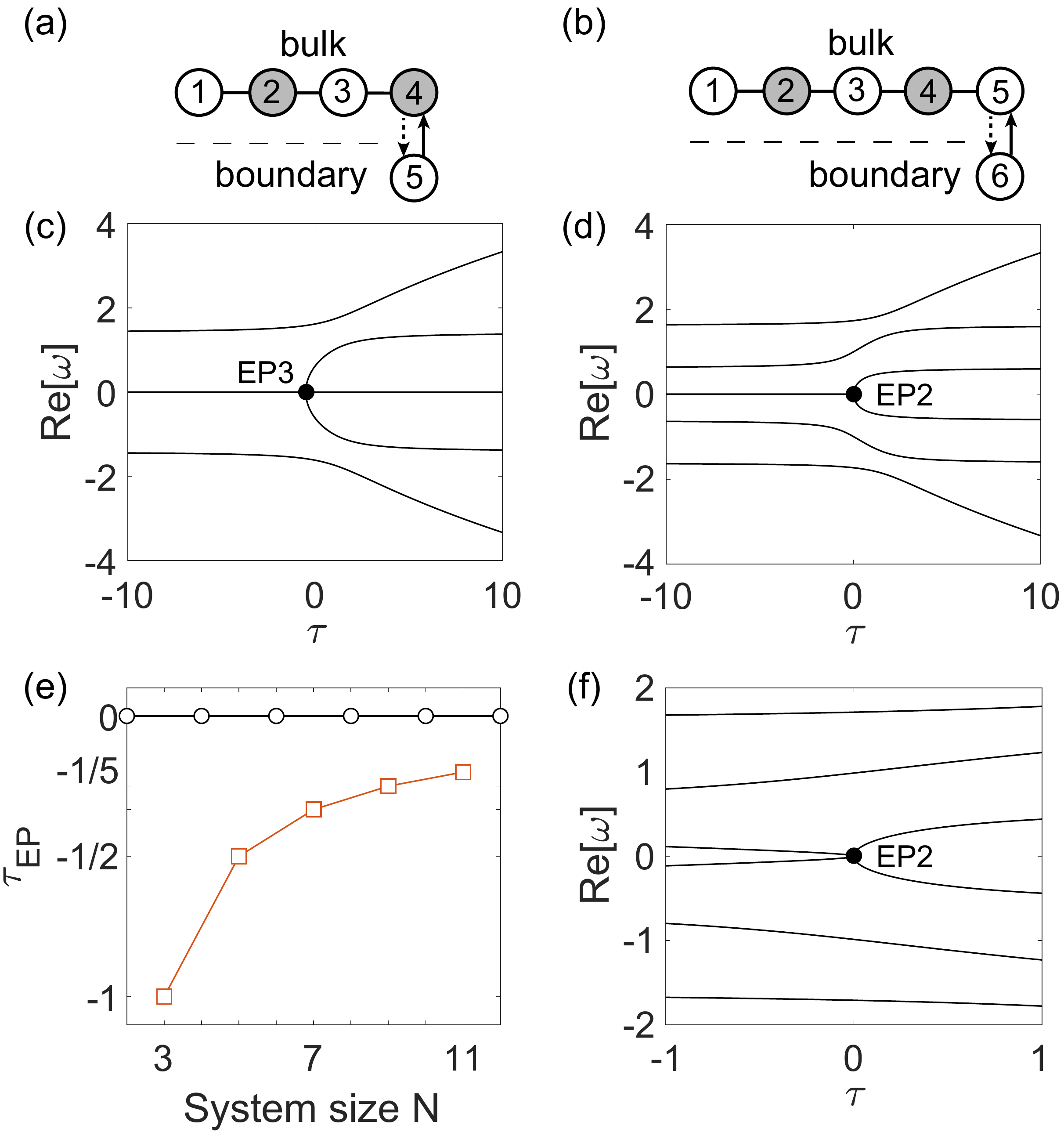}
\caption{Robust and non-robust EPs in 1D cavity arrays with different sizes. (a,b) Schematics showing the 1D lattices with $N=5$ and 6. (c,d) Eigenvalues of $H_N$ as a function of $\tau$ for (a) and (b). (e) Location of the EP as a function of the system size, plotted separately for even (upper) and odd (lower) system sizes. (f) Same as (d) but with $t_n'=e^{i\pi/10}$.
} \label{fig:REP}
\end{figure}

The robust EP reported in Ref. \citenum{REP} exists in the following tight-binding 1D cavity array:
\be
H_{N} = \sum_1^{N-1} (t_n|n\rangle\langle n+1| + t'_n|n+1\rangle\langle n|) + \sum_1^{N}\beta|n\rangle\langle n|.
\ee
Here we consider the first $(N-1)$ cavities as the bulk and the last (i.e., $N$th) cavity as the boundary [see Figs.~\ref{fig:REP}(a,b)]. The on-site energy $\beta$ is taken to be a constant and set to zero across the array, and we start by considering a constant coupling $t_n=t_n'=1$ in the bulk. The bulk couples to the boundary asymmetrically, with $t_{N-1}=1$ and $t'_{N-1}\equiv\tau\neq1$. By tuning $\tau$, a given $H_{N}$ only has a single EP [see Figs.~\ref{fig:REP}(c,d)], and its location (denoted by $\tau_\text{EP}$ below) is plotted in Fig.~\ref{fig:REP}(e) in terms of $\tau$. Here we observe two contrasting behaviors: When $N$ is even, the EPs all occur at $\tau_\text{EP}=0$, which are the robust EPs reported in Ref.~\citenum{REP}; when $N$ is odd, however, they take place at a sequence of $\tau_\text{EP}$'s given by $2/(1-N)$.

To understand this disparity, we first note that the underlying system has non-Hermitian sublattice symmetry \cite{zeromodeLaser}, defined by the anticommutation relation $\{H,C\}=0$. Here $C$ is the sublattice operator given by $C=P_A-P_B$, where $P_{A,B}$ are the projection operators onto the two sublattices of this tight-binding model. These sublattices are defined such that cavities on one sublattice only couple to the ones on the other sublattice, and $C$ in its matrix form is diagonal with alternating elements of 1 and $-1$ here.

One important consequence of sublattice symmetry is a symmetric spectrum, i.e., $\omega_n=-\omega_m$. When the two subscripts $n,m$ are different, the two corresponding wave functions are related by $\psi_n=C\psi_m$, and hence they have the same intensity profile. When these two subscripts are the same, then we immediately find $\omega_m=0$ (i.e., a zero mode \cite{zeromodeLaser}) and $\psi_n$ is a \textit{dark state} with zero amplitude in the sublattice marked by the $-1$ entries in $C$.

Note that this is the \textit{only} wave function of zero modes permitted by the sublattice operator $C$ in a 1D system with two open ends (i.e., not a ring). Therefore, the geometric multiplicity of a degenerate at $\omega=0$ in this case is always 1 independent of its algebraic multiplicity. In other words, the order of an EP with $\omega=0$ equals the number of degenerate modes. For example, the EPs shown in Figs.~\ref{fig:REP}(c,d) are third order (EP3) \cite{Graefe,Loncar,Flatband_PT,Flux,NHFlatband_PRJ,sensingEP3} and second order (EP2), respectively.

\subsection{N-even case}
\label{sec:REP_A}
To show that $\tau_\text{EP}=0$ always holds when the system size is even, we first note that the $N$ eigenvalues of $H_{N}$ form $N/2$ pairs, and the one closest to the origin of the complex energy plane coalesce at $\tau_\text{EP}=0$ and form an EP2, as we have seen in Fig.~\ref{fig:REP}(d). For $N=2$, one can explain this observation directly using
\be
H_2 =
\begin{pmatrix}
0 & 1\\
0 & 0
\end{pmatrix},\label{eq:Jordan2x2}
\ee
which has the Jordan normal form \cite{Heiss} and a coalesced eigenstate $\psi_\text{EP} = [1,\,0]^\text{T}$. The superscript ``T'' here denotes the matrix transpose as usual. To show that this property also holds for larger systems with an even number of cavities, we note the following relation: If $\omega=0$ is an EP2 of $H_N$ and the corresponding wave function has the structure $\psi_\text{EP}=[1, 0,\, \ldots]^\text{T}$ where ``$\dots$" denotes an arbitrary sequence, then $\omega=0$ is also an EP2 of $H_{N+2}$ with wave function $\psi'_\text{EP}=[-1,\, 0,\, \psi_\text{EP}^\text{T}]^\text{T}$. This relation is straightfoward to prove by rewriting $H_{N+2}$ as
\be
H_{N+2} =
\begin{pmatrix}
\sigma_x & V \\
V^\text{T} & H_N
\end{pmatrix},\quad
\sigma_x
=
\begin{pmatrix}
0 & 1\\
1 & 0
\end{pmatrix},
\label{eq:itr1}
\ee
and observing $H_{N+2}\psi'_\text{EP}=0$. Here $\sigma_x$ is the first Pauli matrix and $V$ is an empty $2\times N$ matrix except for $V_{2,1}=1$. To repeat the same analysis for $H_{N+4}$, we just need to let the new $\psi_\text{EP}$ be $-\psi'_\text{EP}=[1, 0,\, -\psi_\text{EP}^\text{T}]^\text{T}$. This coalesced wave function, of course, is the dark state mentioned earlier.

Besides the requirement that the boundary connects only to one cavity in the bulk when $\tau$ is varied, these robust EPs at $\tau=0$ seem to rely only on sublattice symmetry of the system (and more specifically, that of the bulk). To see this property, we replace $\sigma_x$ in Eq.~(\ref{eq:itr1}) by the more general form
\be
\sigma'_x=
\begin{pmatrix}
0 & t_1\\
t'_1 & 0
\end{pmatrix}.
\ee
Our previous proof for $\tau_\text{EP}=0$ still holds by changing $\psi'_\text{EP}$ to $[-1/t'_1, 0,\, \psi_\text{EP}^\text{T}]^\text{T}$ and renormalizing its first element to 1 in the next iteration.  The system still has sublattice symmetry in this case, which holds even when the rest of the bulk couplings in the tight-binding model are random and asymmetric (but nonzero). Therefore, one may attempt to attribute the robustness of these EPs in the presence of coupling disorders to sublattice symmetry. However, as we will show in Sec.~\ref{sec:origin}, robust EPs can persist even when the sublattice symmetry is broken, while they can also be absent in the presence of sublattice symmetry. Based on these findings, we pinpoint the true origin of a robust EP to a restricted bulk zero mode to be discussed in Sec.~\ref{sec:origin}, which can also exist when the bulk has NHPH symmetry.

In this particular 1D lattice, we point out that while both sublattice symmetry and NHPH symmetry are present when $H_{N}$ is real (i.e., with real couplings $t_n,t_n'$), the robust EP here is not supported by NHPH symmetry. NHPH symmetry is specified by $\{H,CT\}=0$ \cite{zeromodeLaser,defectState,NHFlatband_PRJ}, and here $T$ is the time-reversal operator in the form of the complex conjugation. Consequently, the eigenvalues of $H_{N}$ is also symmetric about the imaginary axis in the complex energy plane, i.e., $\omega_n=-\omega_{n'}^*$. Note that unlike in Hermitian systems with particle-hole symmetry, here the complex conjugation is explicit because $\omega_n$'s are complex in general. In the meanwhile, the system also has time-reversal symmetry, i.e.,$[H_N,T]=0$. This is the reason that the eigenvalues of $H_N$ are either real or form complex conjugate pairs, i.e., $\omega_n = \omega_{m'}^*$, similar to PT-symmetric systems. With all its symmetries considered, the eigenvalues of $H_N$ can only be real or imaginary, unless they form quartets ($\omega_n = -\omega_m = \omega_{m'}^*=-\omega_{n'}^*$) which is not the case here. As we mentioned, the robust EP is not supported by NHPH symmetry or time-reversal symmetry here, because it persists when these symmetries are lifted with complex $t_n,t'_n$ [see Fig.~\ref{fig:REP}(f), for example].

\begin{figure}[t]
\includegraphics[clip,width=\linewidth]{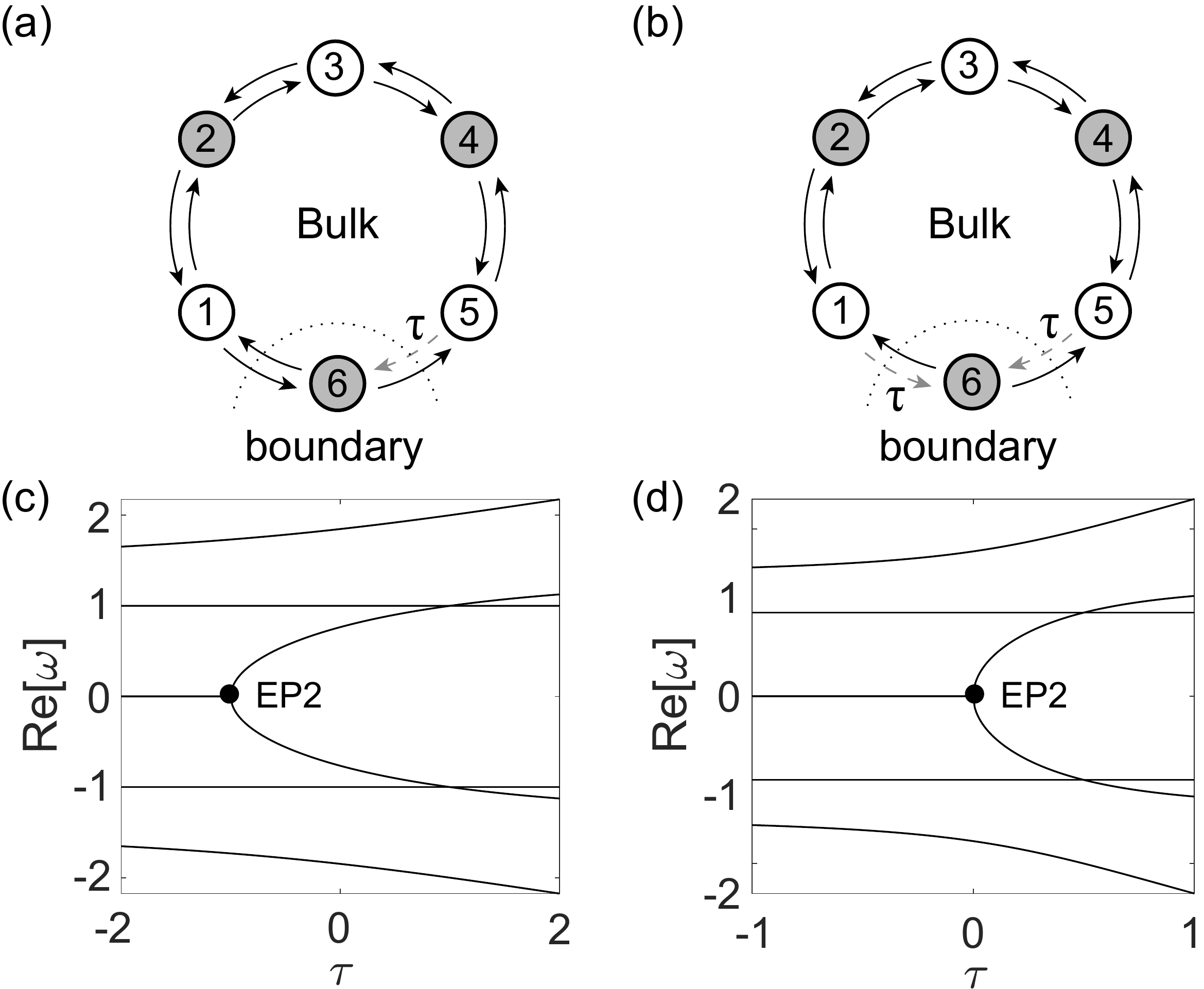}
\caption{Role of the boundary. The boundary (cavity 6) connects to the other side of the bulk (cavity 1) with fixed (a) and tunable (b) couplings. All couplings are set to 1 here other than the dashed ones marked by $\tau$. (c,d) Disappearance and restoration of robust EP at $\tau=0$.
} \label{fig:REP_ring}
\end{figure}

Before we discuss the $N$-odd case, we also emphasize the role of the boundary: if the boundary also connects to another cavity in the bulk, such as in the ring configuration shown in Fig.~\ref{fig:REP_ring}(a), and these additional couplings are fixed and nonzero, then we do not have robust EPs anymore [see Fig.~\ref{fig:REP_ring}(b)]. However, if we do let the additional coupling into the boundary to vanish together with $t'_{N-1}$ [see, for example, Fig.~\ref{fig:REP_ring}(c)], then the robust EP is restored as we show in Fig.~\ref{fig:REP_ring}(d).

\subsection{N-odd case}

When the system size is odd, one of the $N$ eigenvalues of $H_N$ is left alone and must be a zero mode perpetually in the symmetric spectrum $\omega_n=-\omega_m$ warranted by sublattice symmetry. Now when another two originally non-zero modes meet at $\omega=0$, they, together with the perpetual zero mode, form an EP3. This observation is due to sublattice symmetry and holds when coupling disorder is introduced to the bulk. The location $\tau_\text{EP}$ of this EP3, however, is system dependent and hence non-robust as can be seen from Fig.~\ref{fig:REP}(e). Below we show analytically that it is given by $\tau_\text{EP}=2/(1-N)$ when $t_n=t'_n=1$ in the bulk.

We first rewrite $H_N$ in the following basis:
\be
\tilde{\phi}_i
=
\begin{pmatrix}
\phi_i\\
0
\end{pmatrix},\,(i=1,\ldots,N-1),\;\text{and}\;
\tilde{\phi}_{N}
=
\begin{pmatrix}
O\\
1
\end{pmatrix}.\label{eq:newBasis}
\ee
Here $O$ is an empty $(N-1)\times1$ matrix and $\phi_i$'s are the eigenstates of the bulk Hamiltonian $H_B$, i.e., the usual Hermitian tight-binding Hamiltonian given by $H_{N-1}|_{\tau=1}$ without the boundary site $N$. $\phi_i$'s satisfy $\langle\phi_i|\phi_j\rangle=\delta_{ij}$ under the Hermitian inner product, and we denote their eigenvalues by $E_i$'s arranged according to $E_i=-E_{N-i}$. The $i$th and $(N-i)$th eigenstates of $H_B$ are then a pair mapped by sublattice symmetry, and their values in the $n$th cavity satisfy $\phi_i(n)=(-1)^{n+1}\phi_{N-i}(n)$. The resulting Hamiltonian in the new basis reads:
\be
\tilde{H}_{N}
=
\begin{pmatrix}
\bm{E} & \tilde{V} \\
\tau\tilde{V}^\text{T} & 0
\end{pmatrix}. \label{eq:H_newBasis}
\ee
Here $\bm{E}$ is a diagonal $(N-1)\times (N-1)$ matrix with all $E_i$'s, and $\tilde{V}=[\phi_1(N-1),\,\phi_2(N-1),\,\ldots,\,\phi_{N-1}(N-1)]^\text{T}$ is a column vector with the wave functions $\phi_i$'s evaluated in the $(N-1)$th cavity. The characteristic polynomial given by $\text{Det}(\tilde{H}_N-\omega\bm{1})=0$ takes the following form:
\be
\omega\prod_{i=1}^{N-1}(E_i-\omega)+\tau\sum_{i=1}^{N-1}\phi_i^2(N-1)\prod_{j\neq i}^{N-1}(E_j-\omega)=0.\label{eq:Characteristic0}
\ee
Note that the $i$th and $(N-i)$th terms in the summation can be combined to give $-2\omega\tau\phi_i^2(N-1)\prod_{j\neq i,N-i}(E_j-\omega)$, where we have used the property $E_i=-E_{N-i}$ and $\phi_i(N-1)=(-1)^{N}\phi_{N-i}(N-1)$. We immediately find that the characteristic polynomial is proportional to $\omega$. In other words, $\omega=0$ is always an eigenvalue of $\tilde{H}_N$ (and $H_N$), i.e., the perpetual zero mode as we have mentioned.

Because the EP at $\omega=0$ has algebraic multiplicity of 3 (see the discussion just before Sec.~\ref{sec:REP_A}), it indicates that by eliminating this factor of $\omega$ from the characteristic equation, $\omega=0$ is still a solution of the remaining equation at the EP3. Therefore, we can set $\omega=0$ in the remaining characteristic equation and obtain the following equation for $\tau_\text{EP}$ using $E_i=-E_{N-i}$ again:
\be
\prod_{i=1}^{\frac{N-1}{2}}E_i^2+2\tau_\text{EP}\sum_{i=1}^{\frac{N-1}{2}}\phi_i^2(N-1)\prod_{j\neq i}^{\frac{N-1}{2}}E_j^2=0.
\ee
It leads to
\be
\tau_\text{EP}^{-1} = - 2\sum_{i=1}^{\frac{N-1}{2}}\frac{\phi_i^2(N-1)}{E_i^2}.
\ee
Finally, using the well known results in the Hermitian tight-binding model (see, for example, Ref. \citenum{Sols})
\be
\phi_i(p)=\sqrt{\frac{2}{N}}\sin\frac{i\pi p}{N},\quad E_i=2\cos\frac{i\pi}{N+1},
\ee
we find
\be
\tau_\text{EP}^{-1}=\frac{1}{N}\sum_{i=1}^{\frac{N-1}{2}} \tan^2\frac{i\pi}{N+1}=\frac{1-N}{2}.
\ee
In the last step we have used the summation of tangent squares found in Ref.~\citenum{Jolley}, which can be proved using the residue theorem here.

\section{Origin of Robust EP}
\label{sec:origin}


As we have mentioned in the introduction, robust EPs can also exist when sublattice symmetry is broken or absent when the system does have sublattice symmetry. To discuss these different scenarios in a unified framework, we extend Eq.~(\ref{eq:H_newBasis}) to the general case where a robust EP exists. We now refer to the bulk Hamiltonian as $H_B$, which has a set of right and left eigenstates:
\be
H_B\phi^{(R)}_i = E_i\phi^{(R)}_i,\quad\phi^{(L)}_i H_B = \phi^{(L)}_iE_i.
\ee
We restrict our discussion to the case where the bulk itself does not have an EP, which is the case in our discussion of the 1D cavity array in Sec.~\ref{sec:REP} and Ref.~\citenum{REP}. Under this condition, $\phi^{(L)}_i$'s and $\phi^{(R)}_i$'s satisfy the biorthogonal relation $(i,j)\equiv\phi^{(L)}_i\phi^{(R)}_j=\delta_{ij}$, where $i,j\in[1,N-1]$. Note that $\phi^{(L)}_i$ is a row vector while $\phi^{(R)}_j$ is a column vector. By defining
\begin{gather}
\tilde{\phi}_i^{(L)}
=
\begin{pmatrix}
\phi^{(L)}_i \;\;0
\end{pmatrix},\quad
\tilde{\phi}^{(L)}_{N}
=
\begin{pmatrix}
O^\text{T}\;\; 1
\end{pmatrix},\\
\tilde{\phi}_i^{(R)}
=
\begin{pmatrix}
\phi^{(R)}_i\\
0
\end{pmatrix},\quad
\tilde{\phi}^{(R)}_{N}
=
\begin{pmatrix}
O\\
1
\end{pmatrix},\label{eq:WF_HN}
\end{gather}
where $O$ is defined under Eq.~(\ref{eq:newBasis}), we note that $LR=\bm{1}$ where $\bm{1}$ is the identity matrix and
\be
L\equiv
\begin{pmatrix}
\tilde{\phi}^{(L)}_1 \\
\vdots\\
\tilde{\phi}^{(L)}_N \\
\end{pmatrix},\quad
R\equiv\begin{pmatrix}
\tilde{\phi}^{(R)}_1 \,\ldots\,\tilde{\phi}^{(R)}_N
\end{pmatrix}.
\ee
This relation enables us to perform the following basis transformation of $H_N$:
\be
\tilde{H}_N\equiv LH_NR=
\begin{pmatrix}
\bm{E} & \tilde{V} \\
\tau\tilde{U}^\text{T} & 0
\end{pmatrix}. \label{eq:H_newBasis2}
\ee
Here $\bm{E}$ is again a diagonal $(N-1)\times (N-1)$ matrix with all eigenvalues of the bulk. $\tilde{V}=[\phi^{(L)}_1(N-1),\,\phi^{(L)}_2(N-1),\,\ldots,\,\phi^{(L)}_{N-1}(N-1)]^\text{T}$ is a column vector with the left eigenstates evaluated in the $(N-1)$th cavity, and $\tilde{U}$ is similarly defined with $\phi^{(L)}_i$'s replaced by $\phi^{(R)}_i$'s. Note that Eq.~(\ref{eq:H_newBasis2}) applies regardless of whether the system size is even or odd, and we recover Eq.~(\ref{eq:H_newBasis}) when $H_B$ is Hermitian (and $\phi^{(L)}_i=[\phi^{(R)}_i]^\dagger$).

One feature of $\tilde{H}_N$ given by Eq.~(\ref{eq:H_newBasis2}) is that its elements in the last row are all zero when $\tau=0$. Therefore, even though it does not have the Jordan normal form, we know immediately that the first $(N-1)$ eigenvalues of $\tilde{H}_N$ (and $H_N$) are given by just $E_i$'s, i.e., the eigenvalues of the bulk Hamiltonian $H_B$. This observation can be verified directly by multiplying $\tilde{H}_N$ and the corresponding eigenstates, which only have a single non-zero element at the $i$th position. In terms of the eigenstates of $H_N$, they are just $\tilde{\phi}^{(R)}_i\,(i=1,2,\ldots,N-1)$ in Eq.~(\ref{eq:WF_HN}), i.e., the right eigenstates of the bulk plus a vanished amplitude in the boundary cavity. Therefore, in order to to have an EP with $\e=0$ at $\tau=0$, at least one $E_i$ needs to be zero. We denote the number of these bulk zero modes by $n$, and we rearrange the orders of $\tilde{\phi}^{(L)}$'s (and $\tilde{\phi}^{(R)}$'s) such that these $n$ zeros now appear from position $N-n$ to $N-1$ on the diagonal of $\tilde{H}_N$.

In addition, an EP with $\e=0$ at $\tau=0$ also requires at least one of these bulk zero modes to have a finite amplitude in the $(N-1)$th cavity adjacent to the boundary, i.e., $\tilde{V}_i\neq0$. To understand this requirement, let us first consider the case where the corresponding $\tilde{V}_i$'s of all the bulk zero modes are zero. Now we can construct the following wave function
\be
\overline{\phi}_{N}^{(R)} = \tilde{\phi}^{(R)}_{N} - \sum_{i=1}^{N-n-1} \frac{\tilde{V}_i}{E_i}\tilde{\phi}^{(R)}_i,\label{eq:superpose_zeromode}
\ee
without the bulk zero modes. In the basis of $\{\tilde{\phi}^{(R)}_i\}$, $\overline{\phi}_{N}^{(R)}$ is a column vector with the first $(N-n-1)$ elements given by the corresponding $-{\tilde{V}_i}/{E_i}$, followed by $n$ $0$'s and the final element $1$. By acting $\tilde{H}_N$ from Eq.~(\ref{eq:H_newBasis2}) on $\overline{\phi}_{N}^{(R)}$, we find the result to be a column vector of zeros, indicating that $\bar{\phi}_{N}^{(R)}$ is the last eigenvector of $\tilde{H}_N$ and its eigenvalue is zero. Note that $\overline{\phi}_{N}^{(R)}$ is different from all the other zero modes, given by $\{\tilde{\phi}_{i}^{(R)}\}~(i=N-n,\ldots,N-1)$: as mentioned, the amplitudes of these bulk zero modes are zero in the boundary cavity, but $\overline{\phi}_{N}^{(R)}$ has a finite amplitude there due to $\tilde{\phi}^{(R)}_{N}$ in Eq.~(\ref{eq:superpose_zeromode}). Therefore, we conclude that the system does not have an EP at $\tau=0$ in this case. Instead, it has a genuine degeneracy with both algebraic and geometry multiplicity of $(n+1)$.

In contrast, if at least one of these bulk zero modes has a finite $\tilde{V}_i$, we first superpose them modes to form a new set of bulk zero modes (i.e., a basis rotation), such that only $\tilde{V}_{N-1}$ is nonzero among them. Now the action of $\tilde{H}_N$ from Eq.~(\ref{eq:H_newBasis2}) on $\overline{\phi}_{N}^{(R)}$ no longer leads to an empty column vector; it has a single nonzero element in the $(N-1)$th position given by $\tilde{V}_{N-1}$. In other words,
\be
J = \frac{1}{\tilde{V}_{N-1}}\overline{\phi}_{N}^{(R)}
\ee
and $\tilde{\phi}_{N-1}^{(R)}$ form a Jordan chain \cite{Heiss} defined by
\be
H_N J = \tilde{\phi}_{N-1}^{(R)},
\ee
which indicates that $\e=0$ is an EP at $\tau=0$.

So far we have shown that $\e=0$ is an EP at $\tau=0$ if the following two conditions are satisfied: (1) one or more $E_i$'s are zero; and (2) at least one of the corresponding $\tilde{V}_i$'s given by $\phi_i^{(L)}(N-1)$'s is finite. A robust EP further requires that such zero modes in the bulk is robust against couplings disorder, which can exist in the presence of either sublattice symmetry or NHPH symmetry evolved from it as we will show. From these discussions, we conclude that such a \textit{restricted bulk zero mode}, with a finite amplitude in the $(N-1)$th cavity adjacent to the boundary, is the true origin of a robust EP, instead of sublattice symmetry itself in either the bulk or the whole system.

\begin{figure}[b]
\includegraphics[clip,width=\linewidth]{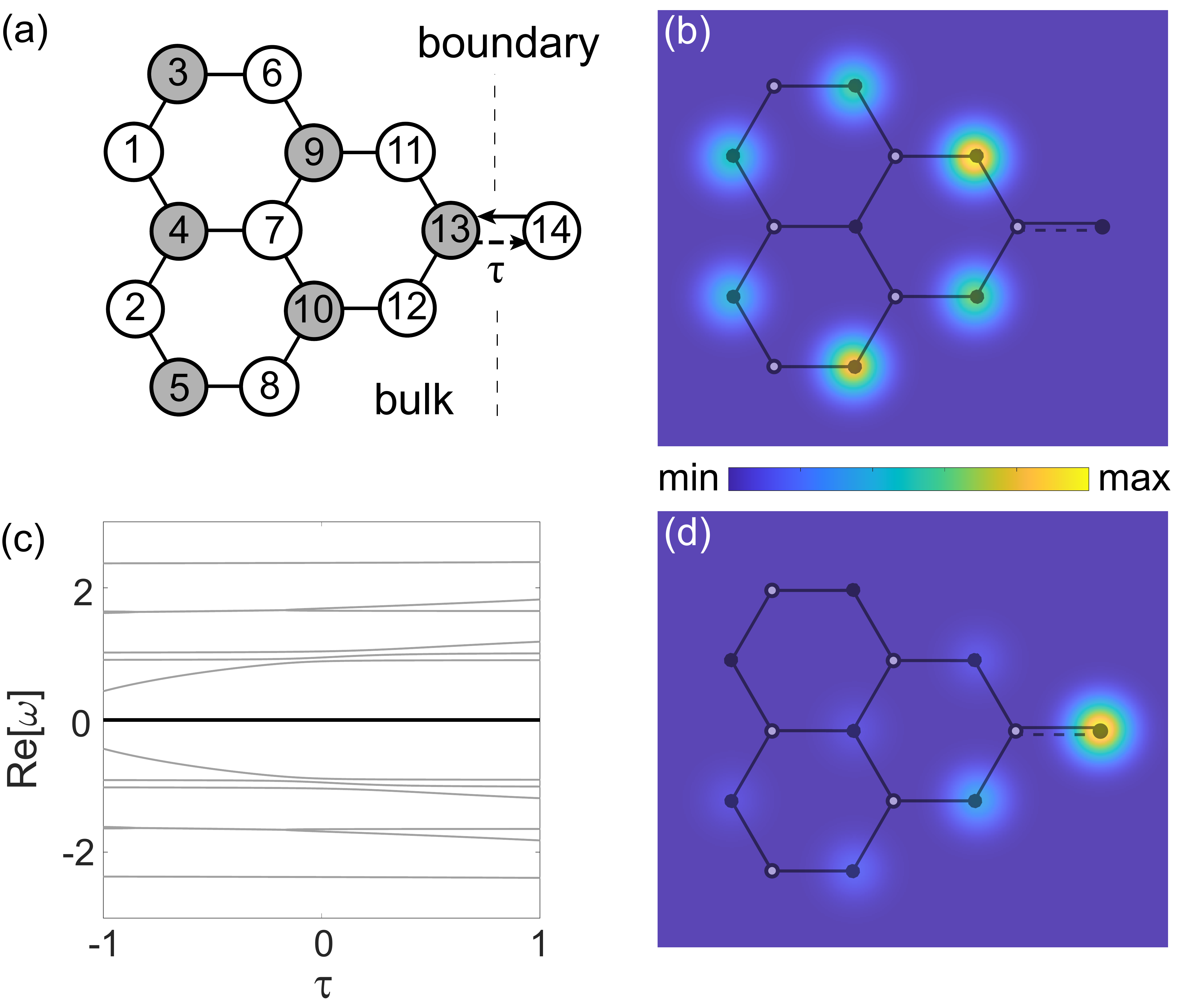}
\caption{Absence of robust EPs with sublattice symmetry. (a) Schematic showing the bulk and the boundary. (b) Bulk zero mode that persists when coupled to the boundary. (c) Real part of all 14 energy eigenvalues. Thick black line shows the two degenerate zero modes, which are independent of $\tau$. (d) New zero mode residing in both the bulk and at the boundary.
} \label{fig:honeycomb}
\end{figure}

To corroborate our conclusion, below we give two illustrative examples. In the first example, the system has sublattice symmetry, and the bulk has a zero mode that however is dark in the $(N-1)$th cavity. Therefore, it does not lead to a robust EP. The system we consider to demonstrate this finding is the tight-binding honeycomb lattice with three rings shown in Fig.~\ref{fig:honeycomb}(a). The bulk Hamiltonian has a single zero mode that is insensitive to coupling disorders thanks to sublattice symmetry, but it has a zero amplitude in the last (i.e., 13th) cavity of the bulk as Fig.~\ref{fig:honeycomb}(b) shows. Therefore, it does not lead to a robust EP: as Fig.~\ref{fig:honeycomb}(c) shows, $\e=0$ is a genuine degeneracy of multiplicity 2 instead (and independent of $\tau$). The wave function of the new zero mode differs from the bulk zero mode by having a finite amplitude in the boundary cavity [see Fig.~\ref{fig:honeycomb}(d)], which indicates the absence of an EP at $\tau=0$ in this case.

\begin{figure}[t]
\includegraphics[clip,width=\linewidth]{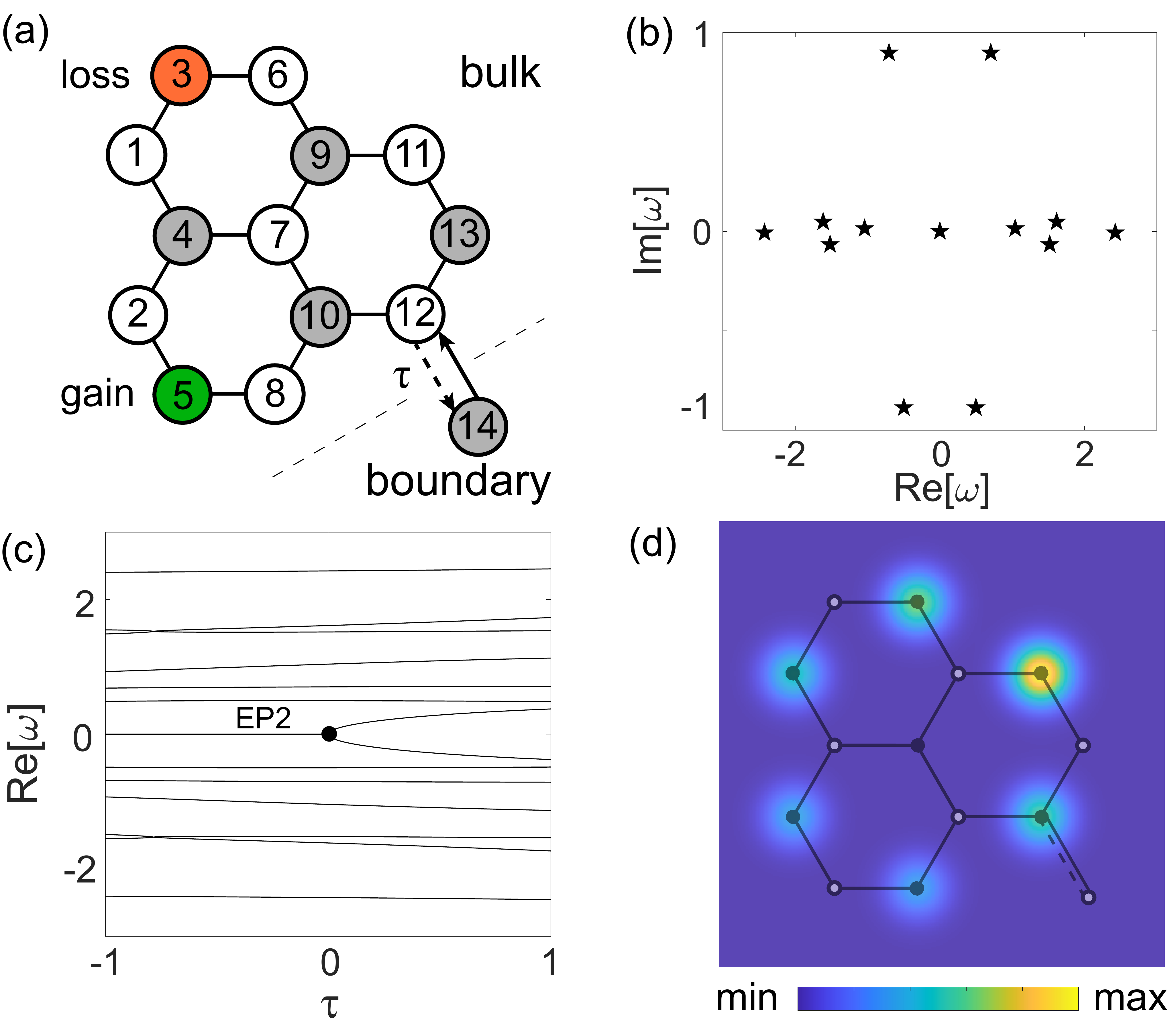}
\caption{Demonstration of a robust EP when sublattice symmetry is broken. (a) Schematic showing the bulk, the boundary, and two cavities with gain and loss. (b) Complex spectrum of the bulk Hamiltonian. (c) Real part of all 14 energy eigenvalues. (d) Wave function of the robust EP at $\tau=0$.
} \label{fig:NHPH}
\end{figure}

In the second example, we show that when sublattice symmetry (and any other non-Hermitian chiral symmetry \cite{NHC_arxiv}) is broken or absent, there can still be a restricted bulk zero mode in the presence of NHPH symmetry and hence also a robust EP. We consider the same honeycomb lattice considered in Fig.~\ref{fig:honeycomb} but with two key differences. First, we couple the bulk to the boundary through cavity 12 [Fig.~\ref{fig:NHPH}(a)], so that the bulk zero mode has a finite amplitude in this cavity and meets the condition that leads to a robust EP. Second, we impose a pair of gain and loss in cavity 3 and 5, and these on-site potentials break the sublattice symmetry of the bulk [see Fig.~\ref{fig:NHPH}(b)] and the whole system.

Nevertheless, the bulk and the whole system now have NHPH symmetry \cite{zeromodeLaser} with the spectral property $\e_\mu = -\e_\nu^*$, as we have mentioned in the introduction and Sec.~\ref{sec:REP}. In addition, the bulk zero mode is not affected by these on-site potentials at $\tau=0$, because it is dark in cavity 3 and 5. As a result, it becomes the wave function of the robust EP2 at $\e=0$ when $\tau=0$. Note that this wave function is different from that shown in Fig.~\ref{fig:honeycomb}(b) due to the coupling disorder in the bulk; it also changes when $\tau\neq0$ here but not in Fig.~\ref{fig:honeycomb}(b). In this example, additional robustness of the EP exists against variations of imaginary on-site potentials, as long as they do not occur in the cavities where the bulk zero mode has a finite amplitude. In fact, we have used two randomly chosen values for the gain and loss strengths in plotting Figs.~\ref{fig:NHPH}(b) and (c), where this additional robustness manifests itself.


\section{Conclusion and Discussion}

In summary, we have analyzed in depth the origin and properties of EPs that are robust against coupling disorder in non-Hermitian tight-binding systems. In the 1D lattice where such a robust EP was found originally, we have shown that it is crucial for this lattice to have an even number of cavities. If instead, one cavity is removed or added to the bulk, the robust EP no longer exists.

By separating the system into the bulk and the boundary, we have established a relation between the energy eigenstates of the bulk Hamiltonian and those of the entire system. This treatment has allowed us to pinpoint the true origin of robust EPs. As we have shown, their existence is not due to the sublattice symmetry of either the bulk or the system, but rather to restricted bulk zero modes that can exist in the presence of either sublattice symmetry or the NHPH symmetry evolved from it when imaginary on-site potentials are introduced.

\begin{figure}[t]
\includegraphics[clip,width=\linewidth]{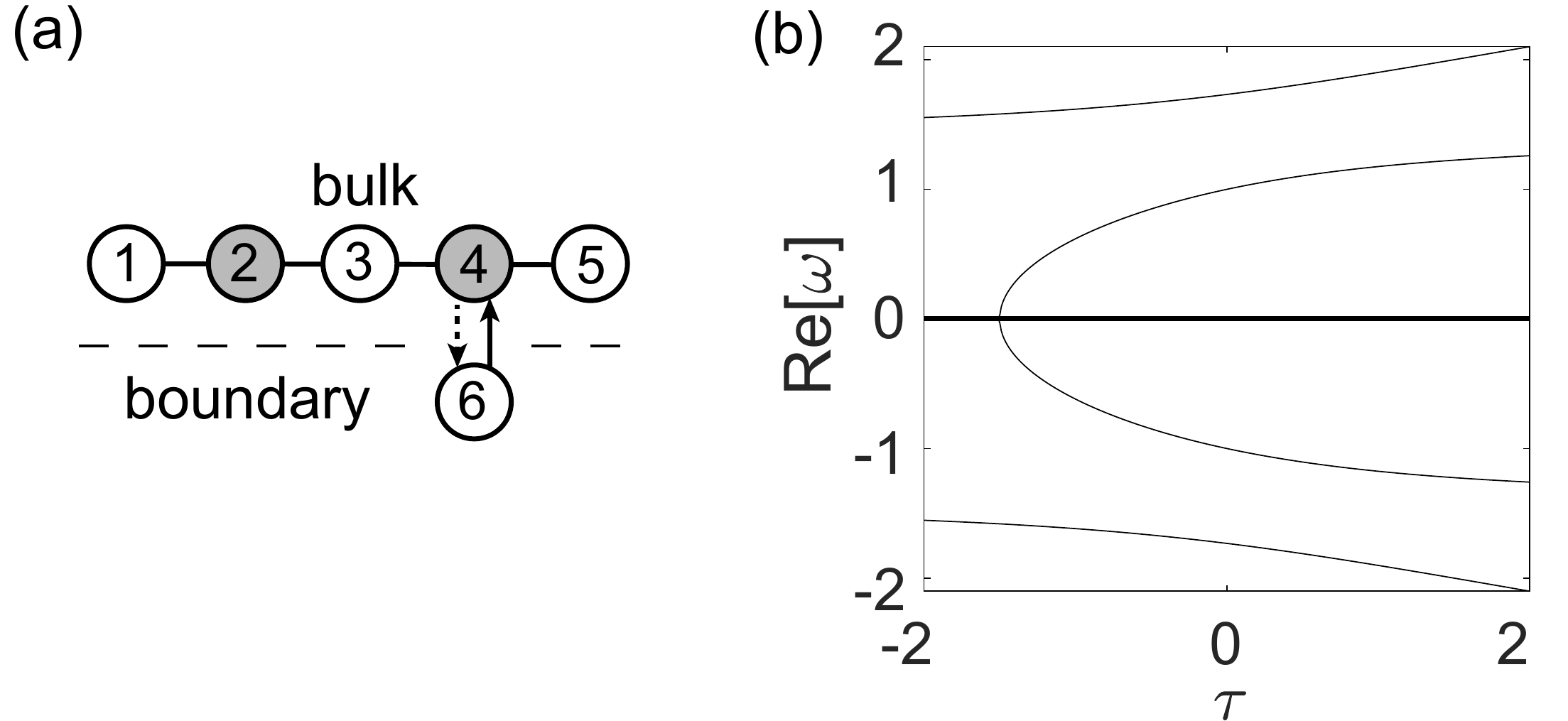}
\caption{Disappearance of the robust EP in Fig.~\ref{fig:REP}(d) with a shifted boundary. (a) Schematic showing the 1D lattices with $N=6$. (b) Real part of the eigenvalues of $H_N$ as a function of $\tau$, showing no robust EPs at $\tau=0$. Thick black line shows the two degenerate eigenvalues with distinct eigenstates, which are independent of $\tau$ and similar to those in Fig.~\ref{fig:honeycomb}(c).
} \label{fig:REP2}
\vspace{-.3cm}
\end{figure}

This observation places an emphasis on the role of the boundary in the analysis of robust EPs. From the two examples we have discussed in Sec.~\ref{sec:origin}, it is striking that whether a bulk zero mode can lead to a robust EP depends on how the bulk is coupled to the boundary. By simply moving the link between the bulk and the boundary to a new location, a robust EP is either destroyed (Fig.~\ref{fig:honeycomb}) or restored (Fig.~\ref{fig:NHPH}). This property holds in other lattices as well, including the 1D lattice we have analyzed in Fig.~\ref{fig:REP}(b): if we now couple the boundary to the $(N-2)$th cavity [see Fig.~\ref{fig:REP2}(a)], i.e., the last but one cavity in the bulk, now the robust EP disappears [see Fig.~\ref{fig:REP2}(b)] since the bulk zero mode (with finite amplitudes only in cavity 1, 3, and 5) no longer has a finite amplitude in the cavity adjacent to the boundary.

This project is supported by the NSF under Grant No. PHY-1847240.


\begin{thebibliography}{99}



\bibitem{Hasan} M. Z. Hasan and C. L. Kane, ``Topological insulators," Rev. Mod. Phys. \textbf{82}, 3045 (2010).
\bibitem{Qi} X.-L. Qi and S.-C. Zhang, ``Topological insulators and superconductors," Rev. Mod. Phys. \textbf{83}, 1057 (2011).
\bibitem{Alicea} J. Alicea, ``New directions in the pursuit of Majorana fermions in solid state systems," Rep. Prog. Phys. \textbf{75}, 076501 (2012).
\bibitem{Beenakker} C. W. J. Beenakker, ``Random-matrix theory of Majorana fermions and topological superconductors," Rev. Mod. Phys. \textbf{87},  1037 (2015).

\bibitem{Sarma_RMP} C. Nayak, S. H. Simon, A. Stern, M. Freedman, and S. Das Sarma, ``Non-Abelian anyons and topological quantum computation," Rev. Mod. Phys. \textbf{80}, 1083 (2008).
\bibitem{Sarma_QI} S. D. Sarma, M. Freedman, and C. Nayak, ``Majorana zero modes and topological quantum computation," npj Quantum Information \textbf{1}, 15001 (2015).
\bibitem{Alicea_PRX} D. Aasen, M. Hell, R. V. Mishmash, A. Higginbotham, J. Danon, M. Leijnse, T. S. Jespersen, J. A. Folk, C. M. Marcus, K. Flensberg, and J. Alicea ``Milestones toward Majorana-based quantum computing," Phys. Rev. X \textbf{6}, 031016 (2016).


\bibitem{NPreview} L. Feng, R. El-Ganainy, and L. Ge, ``Non-Hermitian photonics based on parity-time symmetry,"
Nat. Photon. \textbf{11}, 752--762 (2017).

\bibitem{NPhyreview} R. El-Ganainy, K. G. Makris, M. Khajavikhan, Z. H. Musslimani, S. Rotter, and D. N. Christodoulides, ``Non-Hermitian physics and PT symmetry,"
    Nat. Phys. \textbf{14}, 11--19 (2018).

\bibitem{RMP} V. V. Konotop, J. Yang, and D. A. Zezyulin, ``Nonlinear waves in PT-symmetric systems,"
Rev. Mod. Phys. \textbf{88}, 035002 (2016).

\bibitem{Moiseyev_book} N. Moiseyev, {\it Non-Hermitian Quantum Mechanics} (Cambridge, New York, 2011).

\bibitem{Lu} L. Lu, J. D. Joannopoulos, and M. Soljacic, ``Topological photonics," Nat. Photon. \textbf{8}, 821 (2014).

\bibitem{St-Jean} P. St-Jean et al., ``Lasing in topological edge states of a one-dimensional lattice," Nat. Photon. \textbf{11}, 651--656 (2017).

\bibitem{Bahari} B. Bahari et al., ``Nonreciprocal lasing in topological cavities of arbitrary geometries," Science \textbf{358}, 636--640 (2017).

\bibitem{Bandres} M. A. Bandres et al., ``Topological insulator laser: Experiments," Science \textbf{359}, eaar4005 (2018).

\bibitem{Zhao} H. Zhao et al., ``Topological hybrid silicon microlasers,"
Nat. Commun. \textbf{9}, 981 (2018).

\bibitem{Pan} M. Pan, H. Zhao, P. Miao, S. Longhi, and L. Feng, ``Photonic zero mode in a parity-time symmetric lattice," Nat. Commun. \textbf{9}, 1308 (2018).

\bibitem{Parto} M. Parto et al., ``Complex Edge-State Phase Transitions in 1D Topological Laser Arrays," Phy. Rev. Lett. \textbf{120}, 113901 (2018).

\bibitem{Leykam} D. Leykam, K. Y. Bliokh, C. Huang, Y. D. Chong, and F. Nori, ``Edge Modes, Degeneracies, and Topological Numbers in Non-Hermitian Systems,"
Phys. Rev. Lett. \textbf{118}, 040401 (2017).

\bibitem{Kohmoto} K. Esaki, M. Sato, K. Hasebe, and M. Kohmoto, ``Edge states and topological phases in non-Hermitian systems," Phys. Rev. B \textbf{84}, 205128 (2011).
\bibitem{zeromodeLaser} L. Ge, ``Symmetry-protected zero-mode laser with a tunable spatial profile," Phys. Rev. A \textbf{95}, 023812 (2017).

\bibitem{Bender1} C.~M.~Bender and S.~Boettcher, ``Real spectra in non-Hermitian hamiltonians having $\cal PT$ symmetry," Phys. Rev. Lett. {\bf 80}, 5243 (1998).

\bibitem{PTLaserReview} B. Qi, H.-Z. Chen, L. Ge, P. Berini, and R.-M. Ma, Adv. Opt. Mat. \textbf{7}, 1900694 (2019).

\bibitem{Malzard} S. Malzard, C. Poli, and H. Schomerus, ``Topologically protected defect states in open photonic systems with non-Hermitian charge-conjugation and parity-time symmetry," Phys. Rev. Lett. \textbf{115}, 200402 (2015).

\bibitem{REP} C. Yuce and H. Ramezani, EPL \textbf{126}, 17002 (2019).

\bibitem{EP1} J. Okolowicz, M. Ploszajczak, and I. Rotter,
``Dynamics of quantum systems embedded in a continuum,"
Phys. Rep. {\bf 374}, 271 (2003).

\bibitem{EP2} W. D. Heiss, ``Exceptional points of non-Hermitian operators,"
J. Phys. A: Math. Gen. {\bf 37}, 2455 (2004).

\bibitem{EP3} A. U. Hassan, B. Zhen, M. Soljacic, M. Khajavikhan, and D. N. Christodoulides, ``Dynamically Encircling Exceptional Points: Exact Evolution and Polarization State Conversion," Phys. Rev. Lett. \textbf{118}, 093002 (2017).

\bibitem{EP4} C. Dembowski, H.-D. Gr\"af, H. Harney, A. Heine, W. Heiss, H. Rehfeld, and A. Richter,
``Experimental observation of the topological structure of exceptional points,"
Phys. Rev. Lett. {\bf 86}, 787 (2001).

\bibitem{EP5} S.-B. Lee, J. Yang, S. Moon, S.-Y. Lee, J.-B. Shim, S. Kim, J.-H. Lee, and K. An,
``Observation of an exceptional point in a chaotic optical microcavity,"
Phys. Rev. Lett. {\bf 103}, 134101 (2009).

\bibitem{EP6} M.~Liertzer, L.~Ge, A.~Cerjan, A.~D.~Stone, H.~E.~T\"{u}reci, and S.~Rotter,
``Pump-induced exceptional points in lasers,"
Phys. Rev. Lett. {\bf 108}, 173901 (2012).

\bibitem{EP_CMT} R.~El-Ganainy, M.~Khajavikhan, and L.~Ge, ``Exceptional points and lasing self-termination in photonic molecules," Phys.~Rev.~A \textbf{90}, 013802 (2014).

\bibitem{EP_ring} B. Zhen et al., ``Spawning rings of exceptional points out of Dirac cones," Nature \textbf{525}, 354 (2015).
\bibitem{PT_PRX14} L. Ge and A. D. Stone, ``Parity-Time Symmetry Breaking beyond One Dimension: The Role of Degeneracy," Phys. Rev. X \textbf{4}, 031011 (2014).

\bibitem{NHFlatband_PRJ} L. Ge, ``Non-Hermitian lattices with a flat band and polynomial power increase [Invited]," Photon. Res. \textbf{6}, A10--A17 (2018).


\bibitem{Graefe} G. Demange and E.-M. Graefe, ``Signatures of three coalescing eigenfunctions," J. Phys. A: Math. Theor. \textbf{45}, 025303 (2012).
\bibitem{Loncar} Z. Lin, A. Pick, M. Lon\u{c}ar, and A. W. Rodriguez, ``Enhanced Spontaneous Emission at Third-Order Dirac Exceptional Points in Inverse-Designed Photonic Crystals," Phys. Rev. Lett. \textbf{117}, 107402 (2016).
\bibitem{Flatband_PT} L. Ge, ``Parity-time symmetry in a flat-band system," Phys. Rev. A \textbf{92}, 052103 (2015).
\bibitem{Flux} L. Ge, K. G. Makris, and L. Zhang, ``''Optical fluxes in coupled PT-symmetric photonic structures," Phys. Rev. A \textbf{96}, 023820 (2017).
\bibitem{sensingEP3} H. Hodaei et al., ``Enhanced sensitivity at higher-order exceptional points," Nature \textbf{548}, 187 (2017).

\bibitem{Heiss} W. D. Heiss, ``Exceptional points of non-Hermitian operators,'' J. Phys. A: Math. Gen. \textbf{37}, 2455 (2004).

\bibitem{defectState} B. Qi, L. Zhang, and L. Ge, ``Defect States Emerging from a Non-Hermitian Flatband of Photonic Zero Modes," Phys. Rev. Lett. \textbf{120}, 093901 (2018).
\bibitem{Sols} F. Sols, M. Macucci, U. Ravaioli, and K. Hess, J. Appl. Phys. \textbf{66}, 3892 (1989).
\bibitem{Jolley} L. B. W. Jolley, \textit{Summation of Series}, 2nd rev. ed. (Dover Publications, New York, 2004).

\bibitem{NHC_arxiv} J. D. H. Rivero and L. Ge, ``Chiral symmetry in non-Hermitian systems: product rule, Clifford algebra and
pseudo-chirality," arXiv:1904.02231.

\end{thebibliography}
\end{document}